\documentclass[journal,transmag]{IEEEtran}
\ifCLASSINFOpdf
\else
\fi
\usepackage{graphicx}
\usepackage{subfigure}
\usepackage{amsmath,color,amsthm,amssymb,mathrsfs}
\usepackage{array}
\usepackage{enumerate}
\usepackage{cite}
\usepackage{url}

\begin{document}
%
\title{Dynamic Games for Secure and Resilient Control System Design}


\author{\IEEEauthorblockN{Yunhan Huang\IEEEauthorrefmark{1},
Juntao Chen\IEEEauthorrefmark{1},
Linan Huang\IEEEauthorrefmark{1}, and
Quanyan Zhu\IEEEauthorrefmark{1},}
\IEEEauthorblockA{\IEEEauthorrefmark{1}Department of Electrical and Computer Engineering,
New York University, Brooklyn, NY 11220 USA}
\thanks{Manuscript completed October 16, 2019; submitted October 17, 2019. 
Corresponding author: Yunhan Huang (email: yh.huang@nyu.edu).}}

\markboth{arXiv, October~2019}%
{Shell \MakeLowercase{\textit{et al.}}: Bare Demo of IEEEtran.cls for IEEE Transactions on Magnetics Journals}
%



\IEEEtitleabstractindextext{%
\begin{abstract}
Modern control systems are featured by their hierarchical structure composing of cyber, physical, and human layers. The intricate dependencies among multiple layers and units of modern control systems require an integrated framework to address cross-layer design issues related to security and resilience challenges. To this end, game theory provides a bottom-up modeling paradigm to capture the strategic interactions among multiple components of the complex system and enables a holistic view to understand and design cyber-physical-human control systems.  In this review, we first provide a multi-layer perspective toward increasingly complex and integrated control systems and then introduce several variants of dynamic games for modeling different layers of control systems. We present game-theoretic methods for understanding the fundamental tradeoffs of robustness, security, and resilience and developing a cross-layer approach to enhance the system performance in various adversarial environments. 
This review also includes three quintessential research problems that represent three research directions where  dynamic game approaches can bridge between multiple research areas and make significant contributions to the design of modern control systems. The paper is concluded with a discussion on emerging areas of research that crosscut dynamic games and control systems.
\end{abstract}

\begin{IEEEkeywords}
Game Theory, Dynamic Games, Robustness, Security, Resilience, Cyber-Physical System, Cross-Layer Design, Complex Systems, Sociotechnical Systems.
\end{IEEEkeywords}}

\maketitle

\IEEEdisplaynontitleabstractindextext

%
\IEEEpeerreviewmaketitle

\section{INTRODUCTION}
\begin{figure}
    \centering
    \includegraphics[width=0.8\linewidth]{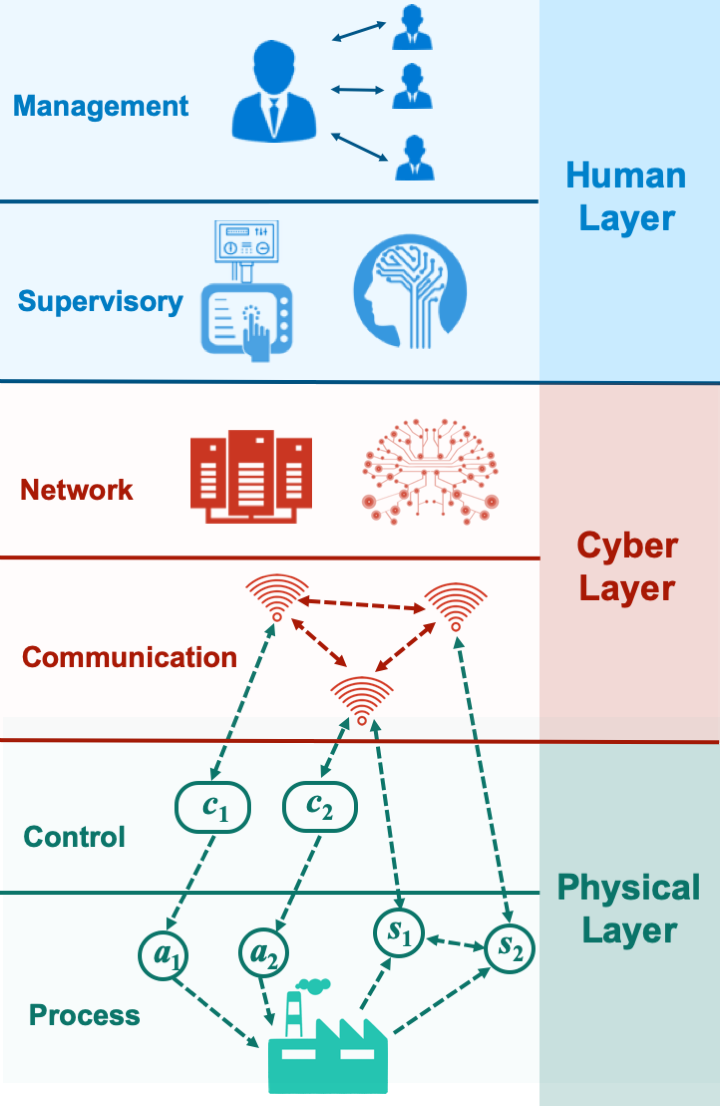}
    \caption{The hierarchical structure of modern control systems is composed of six layers. The physical layer consists of a physical plant with actuators and sensors embedded in it. The control system receives orders, observations, and sends controls to actuators for achieving desired system performance. The communication layer provides wired or wireless data communications that enable advanced monitoring and intelligent control. The network layer allocates network resources for routing and provides interconnections between system units. The supervisory layer serves as the executive brain of the entire system, provides human-machine interactions, and coordinates and manages lower layers through centralized command and control. The management layer resides at the highest echelon. It deals with social and economic issues, such as market regulation, pricing, incentive, and environmental affairs.}
    \label{fig:CrossLayerDesign}
\end{figure}

%
%
%
%
\IEEEPARstart{R}{ecent} advances in information and communications technologies (ICTs) such as the Internet of Things (IoT) and 5G high-speed networks have witnessed increasing connectivity between control systems and cyber networks. The integration between the cyber and physical worlds has made significant advances in many industrial sectors and critical infrastructures, including electric power, manufacturing, and transportation, heralding the fourth industrial revolution that transforms the operation of industrial control systems. To understand and design such systems would require a global and hierarchical perspective toward modern control systems as shown in Fig. \ref{fig:CrossLayerDesign}. The classical view toward control systems consists of sensing, control, and plant dynamics integrated in a feedback loop.

A multitude of control design methods including robust control, adaptive control, and stochastic control have focused on how to deal with uncertainties and physical disturbances \cite{Quanyan2015Magazine}. Modern control systems, due to its exposure to open networks and integration with complex software, require new methodologies that go beyond the classical ones that have focused on the interface between the control layer and the plant at the physical layer. The classical control system is extended by interconnecting it with the cyber and human layers. The cyber layer consists of the communication and networking issues that arise from the communications between sensors and actuators as well as the connectivity among multiple distributed agents. The human layer consists of the supervisory and the management layers that deal with the issues that include coordination, operation, planning, and investment.

As the modern control system design benefits from the growing connectivity, the innate vulnerabilities at the cyber layer and the human layer in modern control systems can bring concomitant threats and hazards from adversaries\cite{pasqualetti2013attack}. Many incidents have been reported as a result of attacker's exploitation of these vulnerabilities \cite{WhiteHouse2013,WSJPowerGrid2019}.  Stuxnet, reported in \cite{Samuel2010,McMillan2010}, is one of the well-known Advanced Persistent Threats (APTs) to control systems that can persist for a long period, behave stealthily, and specifically target industrial control systems by taking advantage of the Supervisory Control And Data Acquisition (SCADA) systems. This type of attacks can also be launched by an insider. One example is the Maroochy water breach incident launched by a disgruntled former employee. 
The attack surface of control systems is exponentially growing. Adversaries can exploit multiple zero-day vulnerabilities and launch unanticipated attacks. One example is the recent hacking of the self-driving vehicles, where the attacker has remotely manipulated, through the cellular connection of the vehicle, various electronic control units, from wiper to brake and engine system \cite{AndyGreenberg2015}. Apart from  self-driving vehicles, many other autonomous systems can face similar threats. Failure to defend against such threats can inflict huge financial losses and fatal damages.
 
The adversarial behaviors at the human and the cyber layers are often hard to anticipate and prepare for. They can cause a significant amount of catastrophic  damage to control systems in terms of their high impact and low effort. The classical approach that regards abnormal behaviors as a result of uncertainties and perturbations to physical plants is insufficient to address these emerging threats. To this end, a new design paradigm is needed to develop frameworks to safeguard the control systems from cyber threats and mitigate the damage that can be caused by attacks. In other words, it is indispensable to consider system properties beyond stability and establish a holistic framework to incorporate the study of robustness, security, and resilience of control systems.

This review aims to present an extensive overview of recent research directions on using game-theoretic approaches to address robust, secure, and resilient design problems of modern control systems. The first objective of this review is to provide a layering perspective toward modern control systems that consist of cyber, physical, and human components across the layers. Game-theoretic methods play an important role in interconnecting different aspects of a control system and providing a holistic and integrated framework to address the cross-layer design of robust, secure, and resilient systems. The second objective of this review is to bridge the classical system design approaches and the modern system design through game-theoretic methods. We can view the secure and resilient control design as an extension of the classical robust control design by integrating multiple game-theoretic frameworks.  Last but not least, the third objective of this work is to introduce the emerging research topics related to game-theoretic methods for secure and resilient control system design. Namely, we present three major application areas including secure and resilient control of heterogeneous autonomous systems,  defensive deception game for industrial control systems, and risk management of cyber-physical networks.

Game theory is not a panacea for all secure and resilient problems in modern control systems. A large number of methods, including event-triggered control, cryptography, detection methods, etc., have provided numerous secure and resilient mechanisms for modern control system design. For those methods, one can refer to \cite{Annaswamy2016}. Game theory have also been applied to many other application scenarios in control systems including multi-agent distributed control\cite{Marden2018,Murray2007}, consensus \cite{bauso2006mechanism}, robust estimation\cite{bacsar2008h}, control of complex systems\cite{Renren2019}. In this review, we focus on game-theoretic methods for robust, secure, and resilient control system design with an emphasis on dynamic games. For game-theoretic security surveys in general Cyber-Physical Systems (CPSs), one can refer to \cite{liang2012game,ramachandran2016dynamic,roy2010survey,wang2016survey,etesami2019dynamic}.

\subsection{The Triplet: Robustness, Security, and Resilience}
Robustness, security, and resilience are three major control system properties for modern control systems. The notion of robustness describes a system's ability to maintain its performance in the presence of regular and singular perturbations \cite{bacsar2008h}, whereas security refers to the system's ability to withstand and be protected from malicious behaviors and unanticipated events \cite{Quanyan2015Magazine}. Robustness and security are two system properties that are achieved offline by foreseeing the perturbations and the attacks before they happen. Thus, these two system properties are classified as pre-event concepts. Despite many endeavors toward designing robust and secure systems, it is impractical and economically inefficient, if it is possible, to achieve perfect robustness and security against all possible perturbations, attacks, and events. This concern calls for the notion of resilience, a post-event concept referring to the system's ability to recover online after adversarial events occur. Hence, resilient control systems have performance guarantees so that even when robustness and security fail under unanticipated attacks and failures, the systems can self-recover from deterioration.

It is imperative to be aware that robustness, security, resilience are three interdependent concepts. These three system properties should be jointly considered in the design of modern control systems. Since a robust control system can withstand a certain range of uncertain parameters and disturbances, the system stays safe under the malicious attacks if the design of security can limit the impact of the malicious attacks within an acceptable range.  Additionally, the design of resilient control systems pivots on the fundamental system tradeoffs between robustness, security, and resilience. Perfect security could be attained by making the system unusable, and likewise, perfect robustness could be reached by considerably degenerating the control performance. The fact that no desirable control systems exhibit perfect robustness or security creates a serious need for resilience. Hence, the three system properties should be jointly designed. It is of vital importance to know, on the one hand, what type of uncertainties or adversarial events need to be considered for enhancing robustness and security, and on the other hand, what uncertainties or malicious events need to be considered for post-event resilience.

\begin{figure}
    \centering
    \includegraphics[width=\linewidth]{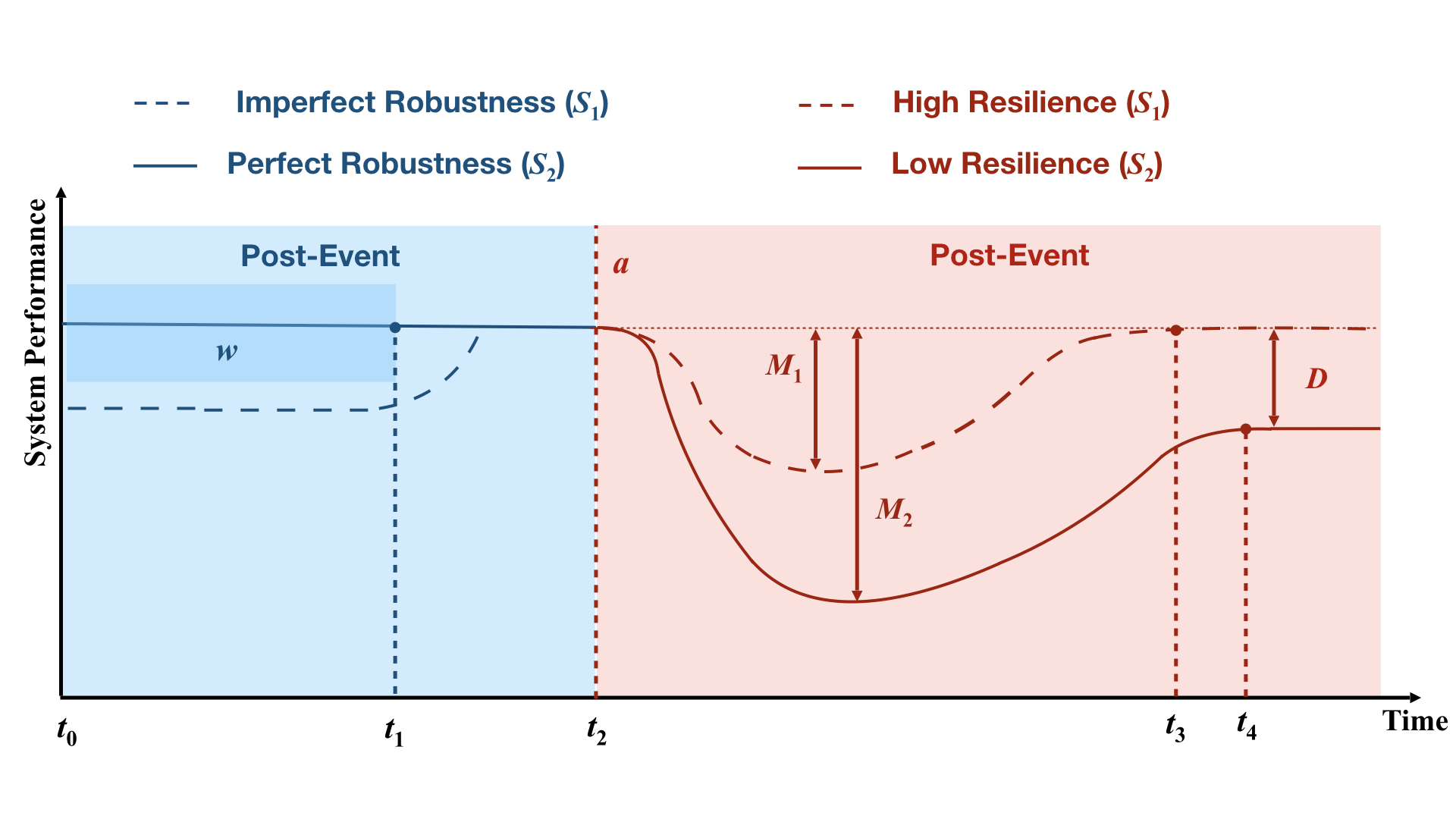}
    \caption{System functionality evolves over time as different events happen. Solid line represents system  $S_1$; dash line represents system $S_2$. Before $t_1$, a known small range of disturbances  $w$ hits the systems. At $t_2$, an attack or rare event $a$ happens. At $t_3$, system $S_1$ finishes full recovery; later at $t_4$, system $S_2$ finishes recovery. System $S_2$ fails to accomplish full recovery and suffer from a steady-state functionality degradation $D$. The maximum functionality degradation of system $S_1$(or resp. $S_2$) induced by the event is denoted by $M_1$(or resp. $M_2$).  }
    \label{fig:ResilienceMetrics}
\end{figure}

Metrics for robustness in control systems have been well established in the literature \cite{bacsar2008h,Kemin1998}. A game-theoretic approach has been introduced to obtain the $H^\infty$ optimal, disturbance-attenuating minimax controllers by viewing the controller as the cost minimizer and the disturbance as the maximizer. Likewise, game-theoretic frameworks have been established to capture the conflict of goals between an attacker who seeks to escalate the damage inflicted on the system and a defender who aims to mitigate it \cite{Manshaei2013}. There is a rich literature on defining metrics for the security \cite{Manshaei2013,Annaswamy2016,Teixeira2015}. However, metrics for security, unlike those for robustness, are problem dependent as the attack model varies and the security design parameters depend on the defense mechanisms such as cryptography, detection, network architecture, and communication protocols. Examples of recent security metrics can be found in  \cite{Dong2010,Quanyan2011ECDC,Quanyan2013CCPS,Yuan2016,Dibaji2019}. Metrics for resilience naturally require a comparison between the pre-event and the post-event performance as resilience is a system property defined as the ability to recover from severe stresses induced by natural disasters or malicious attacks.  Fig. \ref{fig:ResilienceMetrics} illustrates the notion of resilience with respect to an attack that is launched at time $t_1$. Shortly after the attack, the system performance starts to degrade to its maximum degree $M_1$ and $M_2$ for the high-resiliency system ($S_2$) and the low-resiliency system ($S_1$), respectively. Recovery mechanisms are used to restore the system to its original performance or a steady-state degraded performance for system $S_2$ and $S_1$, respectively. A system is said to be more resilient if the system is capable of recovering after an attack with a lower loss of performance and a faster recovery time.

\subsection{Game-Theoretic Methods}

Game theory \cite{Fudenberg1991,Basar1991Dynamic}, in a nutshell, studies the strategic interaction between two or multiple decision-makers, called players, where each player aims to optimize his respective objective function, which depends on the choices of other players in the game. Hence, the optimal decisions of the players are coupled when they aim to achieve the best for themselves. Game theory provides a powerful modeling tool to describe strategic interactions among players. Based on objectives of the players, games can be divided into two categories: zero-sum games and nonzero-sum games. 

A zero-sum game refers to a two-player game where the sum of the two players' objective functions is zero or can be made zero by appropriate positive scaling and/or translation that do not depend on the decision variables of the player.  Zero-sum games are often used to describe conflicting objectives between two players where one player's gain is the other player's loss. Security games often take the form of zero-sum games as in Blotto games \cite{ferdowsi2017colonel} and adversarial machine learning problems \cite{goodfellow2014generative}. 
A non-cooperative game is nonzero-sum if the sum of the players' objective functions cannot be made zero. If each player in a game has only a finite number of alternatives, this game is finite, or a matrix game; otherwise, it is an infinite game. A continuous-kernel game is an infinite game where the action sets of the players are subsets of finite-dimensional vector spaces, and the players' objective functions are continuous with respect to action variables of all players. A game is dynamic when players interact multiple rounds sequentially. A game is of complete information if the structure of the game being played is of common information to all players, including the number of players, the objective functions of the players, the underlying dynamics, the information structure, etc.; it is of incomplete information, otherwise.

The concepts of equilibrium play a vital role in game theory which refers to a joint strategy profile from which no player has a unilateral incentive to change his strategy within the rules of the game. Based on the types of game, we have various notions of equilibrium including, Nash equilibrium, Stackelberg equilibrium, saddle-point equilibrium (SPE), Bayesian equilibrium, etc. They are useful to describe outcomes of different types of interactions among players. For a detailed exposition of basic concepts of equilibrium solutions, we refer the reader to \cite{Basar1991Dynamic,Fudenberg1991}; and for a review of game-theoretic applications to cyber security, we refer readers to \cite{Manshaei2013,zhu2019game,zhu2018game,pawlick2019game}.

Dynamic games are useful to model multi-layer interactions in control systems as the system dynamics evolve, and different components across the players contribute to the path of the dynamics.  For example, the adversary who disrupts the communication channels can create a denial-of-service attack that makes sensor data unavailable and hence leads the plant dynamics toward an unstable trajectory. 
The negligence of a human operator can expose the control system network to malware, which aims to disrupt the normal operations of a nuclear power plant. In dynamic games, the information structure of the game, the form of dynamical systems, and the constraints on the strategy space determine different classes of dynamic game models that are useful to describe a rich class of scenarios of interactions for control systems.  For example, the design of robust control systems has been successfully formulated as a continuous-time differential game between disturbance and controller, which are regarded as two players \cite{bacsar2008h,Kemin1998,Basar1991Dynamic,Quanyan2011ECDC}. The controller seeks to minimize the control cost criterion by choosing a controller that adapts to a given information structure while the disturbance aims to maximize it.

The design of security mechanisms against APT attacks can be viewed as a multi-stage game where an attacker aims to find a path toward the control system network from its initial entry point while the network defender aims to detect and deter the attack from reaching the targeted asset \cite{zhu2018multi,rass2017physical,huang2019adaptive}. If the attacker is prevented from reaching the objective or removed from the system, the system is successfully defended. However, when the network defender fails to safeguard the control system from the attack, the resilience strategies need to be planned to restore the attacked control system to its original operation. Resiliency should be built on the robustness and the security of the system as the post-event resiliency relies on the pre-event designs \cite{xu2016cross,xu2018cross,xu2015secure,xu2017secure,xu2017game}.

Hence, the pre-event secure strategy and the post-event resilient strategy are designed as a result of the game between the defender and the attacker. Despite the fact that security games are structurally different from robust control games and may take different forms depending on attack models \cite{Quanyan2015Magazine,Manshaei2013,Dibaji2019,Pirani2019acc,chen-spie-19,Saurabh2013,Clark2013acc,La2017CDC,Anibal2016,Quanyan2010ACC,Fei2018Automatica,Ugrinovskii2017,Yuchi2017,Xiaobin2019,ferdowsi2017colonel}, both security/resilience and robustness of control systems can be studied using dynamic game frameworks. Thus, dynamic games provide a holistic approach to create an integrated framework to design robust, secure, and resilient control systems by composing different types of games together, as shown in recent literature such as \cite{chen-CDC-16,Dong2010,Quanyan2013CCPS,Pajic2015CDC,Quanyan2015Magazine,Quanyan2011ECDC,Yuan2016,Dibaji2019,chen-spie-19}.

\section{Dynamic Games for Robustness, Security and Resilience}\label{sec: DynamicGameCrossLayer}

Modern control systems primarily consist of six layers: physical, control, communication, network, supervisory,
and management, as illustrated in Fig. \ref{fig:CrossLayerDesign}. Sitting at the bottom is the physical world of the system which serves as a foundation for modern control systems. The physical world of the system can be viewed as an integration of the physical plant to be controlled and the control layer providing control signals based on the feedback. On top of these two layers are the communication layer which establishes wired or wireless communications, and the network layer that allocates resources and manages routing. The communication and network layers constitute the cyber world of the system. Note that in remote control systems, the control layer can be sitting above the cyber layer. Systems containing mainly the cyber layer and the physical layer are called cyber-physical systems. Serving as the brain of the system, the supervisory layer coordinates the cyber and physical layers by designing and sending appropriate commands. Together with the supervisory layer, the management layer interfaces with humans and makes high-level decisions, creating a human-in-the-loop cyber-physical control system.

The design of cyber-physical control system used to be a compartmentalized process, where the cyber system engineers design network protocols and security policies independent from the engineers who design control laws for the underlying physical or chemical processes. This practice, however, is not sufficient to meet the integrated system requirements when the two systems are tightly coupled and strongly interdependent. It is imperative to take into account cyber security when designing control laws for the physical systems, and be aware of the physical impact when designing communications protocols and configuring network devices. 

\subsection{The Cyber-Physical Human System Framework}

\begin{figure*}
\begin{center}
\begin{minipage}[b]{0.44\linewidth}
\begin{center}
     \includegraphics[width=0.96\textwidth]{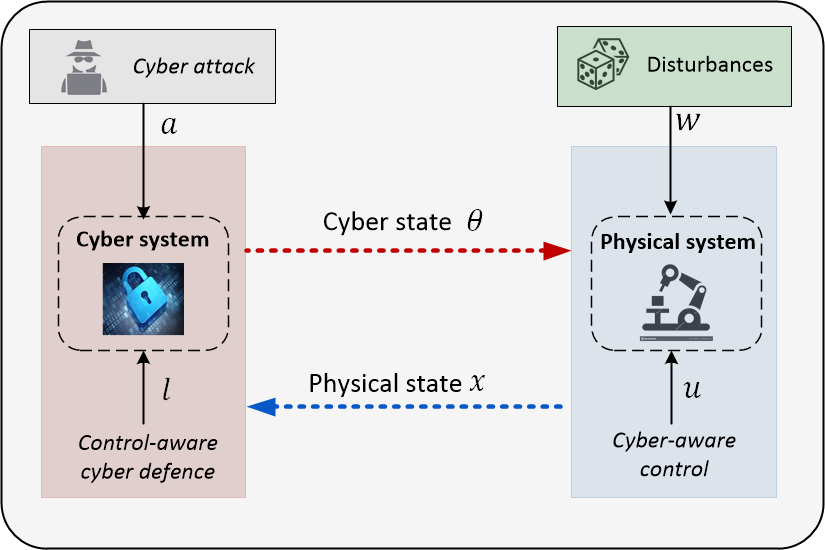} \vspace{-2mm}
  \caption{Illustration of the security-aware resilient control systems.}\label{systemfigure} \vspace{-9mm}
\end{center}
\end{minipage}
\hspace{0.25cm}
\begin{minipage}[b]{0.47\linewidth}
\centering
    \includegraphics[width=0.98\textwidth]{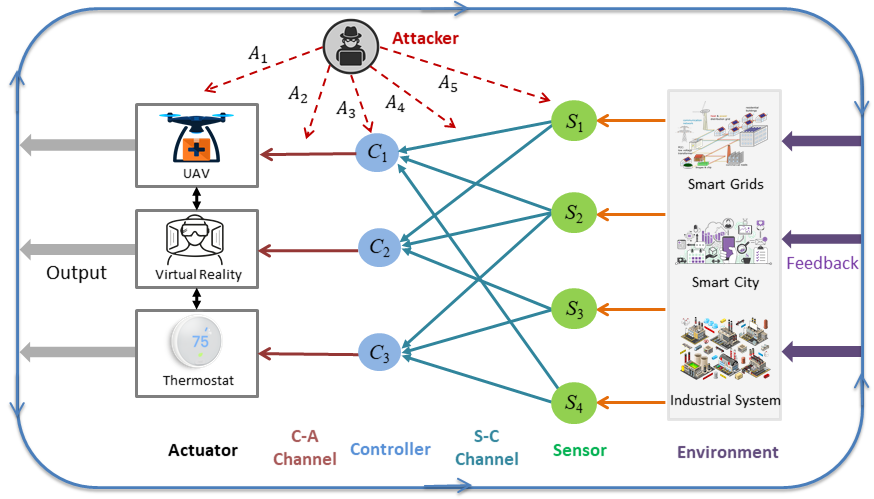}\vspace{-2mm}
   \caption{The attacker can compromise various components in a control system, including the sensors, communications, controllers, and actuators.}\label{attackgraph}\vspace{-9mm}
\end{minipage}
 \end{center}
 \vspace{2mm}
\end{figure*}

The baseline security-aware resilient control systems is illustrated in Fig. \ref{systemfigure}, and can be mathematically described using the following  dynamical system model:
\begin{eqnarray}\label{sys1}
\label{xsystem} \dot{\mathbf{x}}(t) &=& f(t, \mathbf{x}, \mathbf{u}, \mathbf{w}; \theta(t, a, l)), \ \ \mathbf{x}(t_0)=x_0, \\
\label{ysystem} \mathbf{y}(t) &=& h(t, \mathbf{x}, \mathbf{w}; \theta(t,a,l)), \vspace{-2mm}
\end{eqnarray}
where $f,g$ are continuous functions in $(t,\mathbf{x},\mathbf{u},\mathbf{w})$; $\mathbf{x}(t)\in\mathbb{R}^n$ is the state of the physical system; $\mathbf{y}(t) \in \mathbb{R}^m$ is the sensor measurement; $x_0$ is a fixed (known) initial state of the physical plant at starting time $t_0$; $\mathbf{u}(t)\in\mathbb{R}^r$ is the control input; $\mathbf{w}(t)$ models the combined disturbances on the plant and the sensors. The effect of higher layers on the physical layer is encoded in $\theta$ which could be a function of time. The space that $\theta$ lies in is problem-dependent. The evolution of $\theta$ depends on the cyber defense action $l$ and the attacker's action $a$, which could also be functions of time. We use $\theta(t)$ as a shorthand notation in place of $\theta(t, a, l)$ if the pair of actions  $(a, l)$ is fixed.

\subsubsection{Cyber Attack and Defense}
For example, given pair $(a,l)$,  $\theta(t), t\in[0, t_f],$ could be a Markov jump process with right-continuous sample paths, with initial distribution $\pi_0$, and with rate matrix $\lambda=\{\lambda_{ij}\}_{i,j\in\mathcal{S}}$, where $\mathcal{S}:=\{1, 2, \cdots, s\}$ is the state space; $\lambda_{ij} \in\mathbb{R}_{+}$ are the transition rates such that for $i\neq j, \lambda _{ij}\geq0$, and $\lambda_{ii}=1-\sum_{j\neq i}\lambda_{ij}$ for $i\in\mathcal{S}$. 

The framework can be used to capture different types of attacks on control systems, such as the the replay attack \cite{aura1997strategies,miao2013stochastic}, the false data injection attack \cite{bobba2010detecting}, and the sensor attack \cite{mitra2019byzantine}.
\begin{enumerate}
    \item In the replay attack, the attacker can record sensor measurements, choose the replay window size $T_R>0$, and decide whether to send the original or modified sensor outputs at each time step. Let $\theta=\theta_1$  denote the state of the cyber state where there is no attack, and the control system is in a healthy state. Let $\theta=\theta_2$  denote the state where an attack has been successfully launched in the cyber layer, and the control system is compromised. The replay attack can be captured by letting $h(t, \mathbf{x}, \mathbf{w}; \theta_2) = \mathbf{y}({t-T_R})$ in (\ref{ysystem}), stating that the past measurements $\mathbf{y}({t-T_R})$ are taken as the current ones $\mathbf{y}(t)$.
    
    \item In the false data injection attack where the attacker injects data to a subset of sensors, the model (\ref{ysystem}) can be used to capture the attack by letting $h(t, \mathbf{x}, \mathbf{w}; \theta_2) = h(t, \mathbf{x}, \mathbf{w}; \theta_1) + \mathbf{y^a}(t)$, where $\mathbf{y^a}(t)$ is the data value injected by the attacker. In cases where an attacker can cause disruptions to the system operation, for example, by opening a valve in water distribution systems \cite{WaterRoadmap}, or turning on a circuit breaker in electric power systems \cite{EnergyRoadmap}, the dynamics  of the control system will be changed, and they can be captured in (\ref{xsystem}) by specifying the changed post-attack dynamics. Fig. \ref{attackgraph} illustrates the vulnerabilities of control systems to multiple potential attacks, where controller-actuator (C-A) channel and sensor-controller (S-C) channel are vulnerable to cyber attacks. $A_5$ represents direct sensor attacks which can disable a set of sensors. $A_3$ and $A_1$ represent the DoS attacks that prevent controllers from receiving sensor measurements or actuators from receiving control signals. $A_4$ and $A_2$ represent data injections attacks, where the false information $\mathbf{\tilde{y}}\neq \mathbf{y}$ and $\mathbf{\tilde{u}}\neq \mathbf{u}$ is sent from sensors and controllers. 
    \item In sensor attacks, $\theta(t)$ can describe the set of sensors whose signals cannot be received by the control center due to network failure or sensor failure cased by denial of service (DoS) attacks. Each sensor has two states: functioning normally or not. If the number of sensors in the physical plant is $N$, then $\theta(t)\in \mathcal{S}$ and $\mathcal{S}=\{1,...,2^N\}$. At time $t$, the cyber attack action $a(t)$ will be to choose a set of sensors to attack and the cyber defense move $l(t)$ will be to recover a chosen set of sensors. Then, $\{\theta(t)\}_{t\in [0,t_f]}$ becomes a controlled Markov jump process with transition rate $\lambda_{i,j}(a,l),i,j\in\mathcal{S}$. In this case, the system dynamics is considered to be independent from $\theta$, i.e., $f(t,\mathbf{x},\mathbf{u},\mathbf{w};\theta(t,a,l))=f(t,\mathbf{x},\mathbf{u},\mathbf{w})$. The output $y$ is captured by (\ref{ysystem}). For linear system models, we have $\mathbf{y}=C(\theta(t,a,l)) \mathbf{x}$ where matrix $C$ is a function of $\theta(t)$ decided by the set of sensors that function normally. With different $\theta$, the system designer needs to adapt different schemes to do filtering and control.
\end{enumerate}

The costs of launching attacks and executing defenses are captured by $C_A(a,l)$ and $C_D(a,l)$, respectively. The attacker aims to minimize the cost of attacking and deteriorating system performance. Adversely, the system operator aims to minimize the cost of defending and maintaining system performance. If $C_A(a,l)+C_D(a,l)=0$, the attack-and-defense problem is a zero-sum stochastic game \cite{shapley1953stochastic} with a Markov decision process sitting behind. In general, we have $C_A(a,l)+C_D(a,l)\neq0$. The costs $C_A$ and $C_D$ depend on the attacker's and the system's actions and the system performance encoded in $\mathbf{x}$ while the evolution of $\mathbf{x}$ is dependent on $\mathbf{u}$ and $\theta$. Thus, the security and resilience design in the cyber layer is coupled with the system dynamics in the physical plant which should be jointly considered.

\subsubsection{Robustness and Resilience in the Physical Layer}

Given the cyber security strategy pair $(a,l)$, the goal of robust and resilient control is to design a controller that minimizes the performance loss due to the attack, which is measured by the shaded area in Fig. 5. This design problem can be captured by an $H^\infty$ control problem with the performance index given by the expected cost over the statistics of $\theta$:
\begin{equation}\label{PSG}
\begin{aligned}
\ \ \ \ \inf_\mathbf{u} \sup_\mathbf{w}  \ J_P(\mathbf{u}, \mathbf{w}):=&\mathbb{E}_\theta \{ q_f(\mathbf{x}(t_f); \theta(t_f))+\\
& \displaystyle\int_{t_0}^{t_f} g(t, \mathbf{x}(t), \mathbf{u}(t), \mathbf{w}(t); \theta(t))dt \} ,\\
\end{aligned}
\end{equation}
where $q_f$ is continuous in $\mathbf{x}$, and $g$ is jointly continuous in $(t, x, u, w)$. In the infinite-horizon case, $q_f$ is dropped out, and $t_f\rightarrow \infty$. \textcolor{black}{The $H^\infty$-optimal control problem in the time domain is in fact a minimax optimization problem, and hence a zero-sum differential game, where the controller $\mathbf{u}$ can be viewed as the minimizing player and the disturbance $\mathbf{w}$ as the maximizing player \cite{bacsar2008h,Basar1991Dynamic}.} The game (\ref{PSG}) is referred to as the physical system game (PSG), and its solution is characterized by SPE. This framework enables the design of {\it robustness} and {\it resilience} within the same model, and takes into account the security vulnerabilities from the cyber systems. A large number of papers \cite{Saurabh2013,chen-CDC-16,Clark2013acc,ferdowsi2017colonel,La2017CDC,Anibal2016,Quanyan2013CCPS,Quanyan2010ACC} has adopted the idea of deploying dynamic games for the security and resilience of modern control systems with interdependent cyber and physical layers. 

\subsubsection{Cyber-Physical Co-Design and Tradeoffs among Robustness, Security and Resilience}

The cyber-physical nature of modern control systems requires a cross-layer approach for designing secure and resilient systems.  Independent designs of the cyber and the physical layers of the system without knowing their interdependencies  often lead to unintended performance degradation. 
Thus, a co-design process that coordinates between cyber and physical layers of the system is pivotal for the control system. 
As illustrated in Fig. \ref{systemfigure},  the two design processes can be composed together and reach an iterative process for cyber-physical co-design. The resilient control design pair $(\mathbf{u}, \mathbf{w})$ will be used by the cyber system for the design of defense strategy pair $(a, l)$, and likewise, the strategy pair $(a, l)$ is also used by the physical system for the design of the control pair $(\mathbf{u},\mathbf{w})$. The coupled system leads to a {\it holistic} design framework that enables robust, secure and resilient design of infrastructural systems.  The fundamental tradeoffs between robustness, security, and resilience can be quantitatively analyzed and designed:

\begin{itemize}
\item {\it Tradeoff between Robustness and Resilience}: 
Perfect robustness of control systems is not achievable for all types of disturbances and events. However, resilience can be used as a post-event measure to recover the system from the impact of the disturbances and events that are not accounted for in the model. This tradeoff is captured by PSG for the security-aware resilient system design. 
\item {\it Tradeoff between Security and Resilience}: 
Perfect security that is capable of defending against all types of attacks is not realistic for control systems. However, the resilient cyber systems can be designed to quickly bring a compromised state  to their normal operations. This tradeoff is captured by the cyber system game (CSG) for the impact-aware proactive cyber defense.
\item {\it Tradeoff between Robustness and Security}:
The two tradeoffs above lead to a relation between the robustness of the physical system and the security of the cyber system. The high demand for robustness requires a strong level of security.  Given limited resources, they cannot be achieved at the same time. This tradeoff is captured by the coupled PSG and CSG frameworks. 
\end{itemize}

\subsubsection{Human Factors of Control Systems}\label{HumanOnCPS}

The human factors arise from the interactions between the control systems with the supervisory layer and the management layer. The supervisory layer provides human-machine interactions and coordinates and manages lower layers through  centralized command and control as illustrated in Fig.\ref{fig:CrossLayerDesign}.  The behaviors of human designers and human operators are often less predictable and difficult to describe.  They are often viewed as the weakest link in the control system. Attackers can leverage human vulnerabilities to enter and penetrate the multi-layer control system network. 

For example, in the Stuxnet attack \cite{McMillan2010,Samuel2010}, the maintenance engineer connected an infected USB to his maintenance laptop from which the malware comes into the private network and causes SCADA infection. And in the Maroochy breach \cite{Jill2007}, a former employee installed a SCADA configuration program on his own laptop and took control of 150 sewage pumping stations resulting in severe environmental damage. 

The human factors have been studied extensively in the game theory literature with the objective to describe the cognitive, memory, computational, and psychological aspects of the human decision-making process \cite{dhami2016foundations,sanders1998human}. One important area of research is the bounded rationality which captures the behavioral and imperfect decision-making of humans. Several elements in the game-theoretic framework in CSG and PSG can be revised to capture human errors in decision making due to limited memory, attention, or reasoning power. For example, by leveraging the concept of hyperbolic discounting, we can model the time-inconsistent human preferences, which have been demonstrated \cite{thaler1981some} to show that human makes irrational choices at different times. Prospect theory \cite{tversky1992advances,kahneman2013prospect} incorporates loss-aversion in human decisions and differentiates the perception of losses from the utility of the gains. It can be used to extend the risk-neutral decision-making in CSG and PSG to their risk-averse counterparts to understand the consequence of the cognitive bias in the decision-making. 

Attention is another important human factor that can be incorporated in the decision making to capture the limited cognition of the human when they make online decisions \cite{sims2010rational}. Authors in \cite{chen_2019_TIFS} have presented an attention-constrained risk analysis model to assess risks over interdependent risk networks.  
The management layer at the highest echelon deals with social and economic issues, such as market regulation, pricing, and incentives. Players in this layer deal with socio-economic issues involving many stakeholders related to the control systems and make service-level contracts to reduce cyber-physical risks. For example, cyber insurance is an example of financial products to transfer the risk from the control system and mitigate the losses due to cyber threats. Authors in \cite{zhang2017bi,hayel2015attack} have designed incentive-compatible attack-aware cyber insurance policies to maximize the social welfare and alleviate the impact of moral hazard. In \cite{chen2017security}, the authors have designed service contracts for security services in the cloud-enabled autonomous systems.  

As modern control system scales to billions of connected devices and is increasingly complex, it is not always possible for an entity to own and manage all cyber and physical components of the control system.  For example,  in cloud-enabled systems \cite{pawlick_TIFS_18,chen2016optimal,xu2015secure}, smart homeowners  use the services provided by the cloud service provider (SP) who fuses data and optimizes control decisions for real-time systems. Small business owners may not own the sensors but subscribe to service providers (SPs) who collect data that allow users to develop control system applications instead of making a costly investment in their own sensing infrastructure \cite{craciunas2010information}.   

The decentralized ownership and the provision of control system services provide an effective sharing and utilization of the resources of computational, communication, and sensing infrastructures.  In this paradigm, the SP owns the cyber infrastructure and determines defense strategy $l$ while the user owns the physical infrastructure and designs control $\mathbf{u}$. However, a user cannot directly control or manage the security risk. If the SP is negligent in assuring cybersecurity, then users who rely on these services will be subject to high-security risks. It is essential to develop appropriate incentive-compatible service mechanisms for the SPs to offer high quality of services (QoS) while making efforts to mitigate security risks at the service level of control systems. SPs should be incentivized to deploy adequate security mechanisms to ensure the reliability and the dependability of the services for control system users.  It not only enables the  implementation and investment of security but also prevents the cyber risks from further propagating  at the socio-economic scale.

Challenges in the design of the cyber-physical contract come from incomplete information and adversarial behaviors. The incomplete information can arise from the hidden type and the hidden action of the SP. In \cite{chen-tifs-17,chen2016optimal}, the authors have used contract design principles to  develop a holistic incentive-compatible and cost-efficient security-aware service mechanism for real-time operation of cloud-enabled Internet of Controlled Things (IoCTs) under APTs.

\section{Recent Advances}


With the hierarchical perspective toward robust, secure, and resilient control systems, this section aims to introduce several recent dynamic applications to cross-layer control design in adversarial environments. Game-theoretic approaches have been natural frameworks to model conflicts between an attacker and a defender in various scenarios at the communication and networking layers including intrusion detection \cite{nissim2015detection,Quanyan2010ACC,farhang2014dynamic,ghafouri2016optimal,sayin2018game,zhu2010heterogeneous},  jamming and eavesdropping \cite{gupta2010optimal,li2015jamming,mallik2000analysis,mukherjee2012jamming}, and honeypot/deception \cite{pawlick2018modeling,huang2019deceptive,zheng2012dynamic,zhu2012deceptive,DBLPhln}. These approaches use different game models \cite{Basar1991Dynamic,Fudenberg1991} such as zero-sum game, stochastic game, repeated game, differential game, Stackelberg game, etc., to handle different type of attacks including jamming attack \cite{gupta2010optimal,li2015jamming,mallik2000analysis}, eavesdropping \cite{mallik2000analysis}, DoS attack\cite{Yuan2016,Yuchi2017}, replay attack\cite{aura1997strategies,miao2013stochastic}, zero dynamics attack\cite{Teixeira2015,hoehn2016detection}, data injection attack\cite{bobba2010detecting,ghafouri2016optimal}, covert attack\cite{laszka2013mitigation,hoehn2016detection} and cyber epidemic attack\cite{8779673} etc.

Apart from those at the cyber layers, game theory has also successfully addressed risk management \cite{chen-dgaa,zhang2017bi} and security investment problems \cite{chen_2019_TIFS,cavusoglu2008decision} at the human layers and the problem of adversarial consensus \cite{el2016zero,bauso2006mechanism,pirani2019design} and resilient infrastructures \cite{chen_game_book,chen_TIFS_dynamic_game_19,huang2017large,huang2018factored,chen2016game,semiNuclear} at the control layers.

This section presents three quintessential research problems that represent three distinct directions where dynamic game approaches can be useful to bridge between multiple research areas and make significant contributions to the design of modern control systems. The first one leverages a moving-horizon dynamic game technique to secure the heterogeneous autonomous vehicles and enable self-healing after attacks. The second research direction investigates an impact-aware multi-stage cyber deception game where the defender proactively deters the stealthy APT attacks from reaching the critical asset of industrial control systems. Adversarial and defensive deceptions across the entire intrusion process introduce the games of incomplete information, thus both players need to make judicious actions under persistent uncertainty. The third direction focuses on the risk management of networked systems by incentivizing agents to comply with security guidelines with maximum effort.

\subsection{Games for Secure Control of Heterogeneous Autonomous Systems}\label{subsec:HeterAutoSys}

Multi-layer networks or network-of-networks have been seen in a number of critical applications, such as energy and water networks \cite{kurian-17}, power and transportation networks\cite{huang-gamesec-17}, and multi-layer robotic systems \cite{chen-CDC-16}.  Traditional defensive mechanisms for single networked systems are no longer sufficient for this network-of-networks paradigm. To design secure and resilient control strategies for the multi-layer autonomous systems, it is imperative to analyze three types of games resulting from the strategic interactions: i) interactions among agents in individual network layers, ii) interactions between agents from different layers, and iii) interactions between agents and adversaries. 
To address this challenge, the authors in \cite{chen-TCNS-19-games,chen-spie-19} have proposed a \textit{games-in-games} model which is able to understand the network performance, heterogeneous agents' functionalities, and the network operators' decisions holistically. 

\begin{figure}
    \centering
    \includegraphics[width=1\columnwidth]{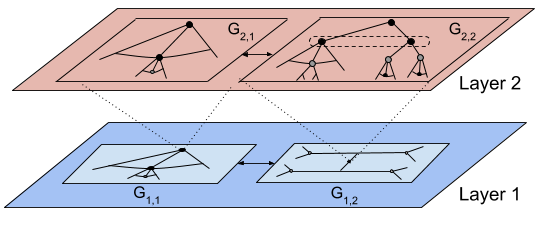}
    \caption{Games-in-games framework for secure and resilient control of multi-layer multi-agent systems. The control of each agent considers the behaviors of the agents at the same layer and the ones at the other layer. Furthermore, the agents also learn and respond to the unanticipated events, such as natural disruptions and adversarial attacks, at each step of decision making. }
    \label{games-in-games}\vspace{2mm}
\end{figure}

\begin{figure*}[!t]
  \centering
      \subfigure[Secure configuration of robotic networks]{
    \includegraphics[width=0.8\columnwidth]{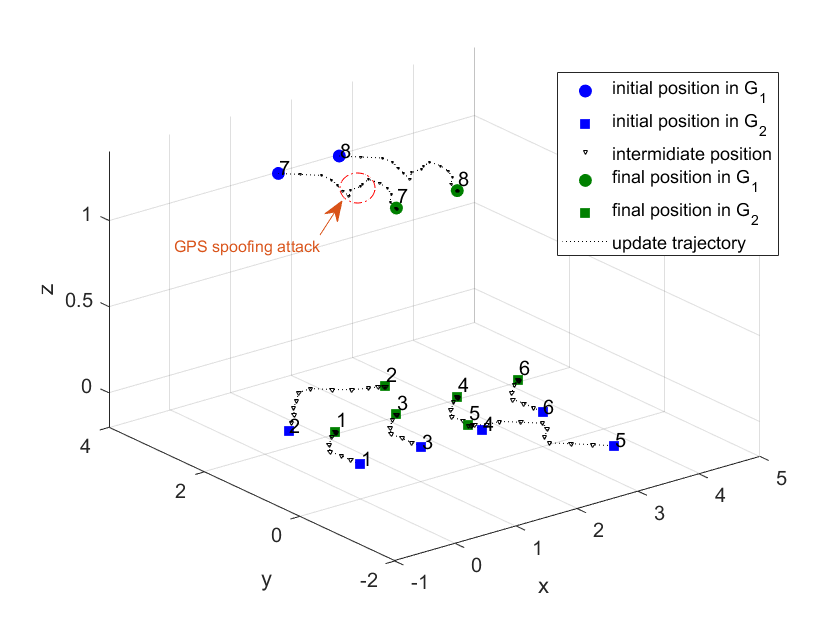}}
      \subfigure[Network connectivity]{
    \includegraphics[width=0.8\columnwidth]{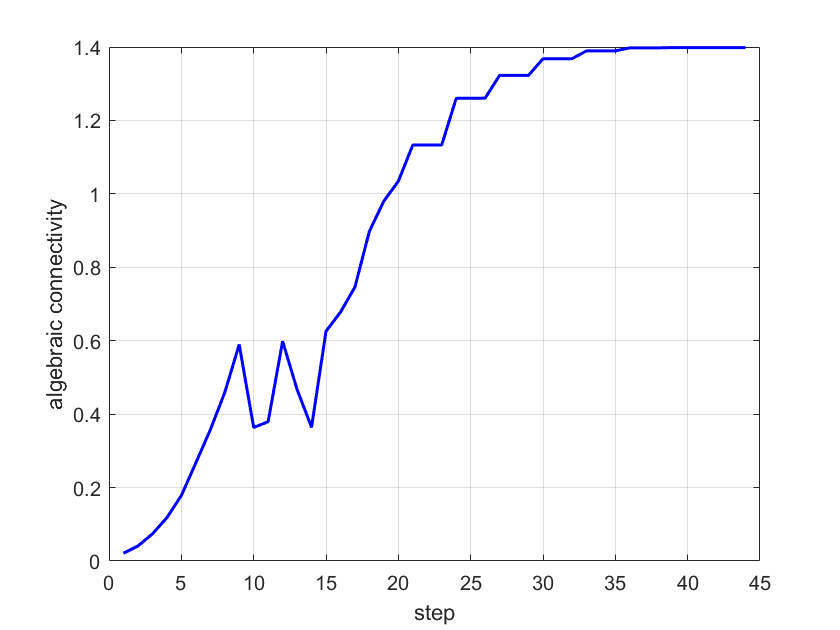}\label{spoofing_connectivity}}
  \caption[]{ (a) shows the dynamic and secure configuration of a two-layer robotic network. The GPS spoofing attack introduced at time step 9 and  lasted for 5 steps. (b) shows the corresponding network connectivity.}
  \label{spoofing_case}
\end{figure*}

For clarity, a pictorial illustration of the games-in-games framework is shown in Fig. \ref{games-in-games}. The system composes two layers of networks. In each sub-layer, agents make decisions based on not only the behaviors of the agents at the same layer but the ones at the other layer. At each step of decision making, the agents also learn and respond to the unanticipated events in an agile fashion, such as natural disruptions and adversarial attacks. Leveraging the framework, one can compose the distinct games together to obtain the Gestalt Nash equilibrium (GNE) \cite{pawlick_TIFS_18,chen_2019_TIFS}. The GNE describes an equilibrium solution concept at which no agent has incentives to deviate away from not only each modular game, which captures the local agent-agent level interactions, but also the integrated game, which considers the global system-system level interactions.

Based on \cite{chen-TCNS-19-games}, we next present an example of controlling two-layer mobile autonomous systems in the adversarial environment. There are three players in the game: two network operators and an attacker. The focused objective in \cite{chen-TCNS-19-games} is the algebraic connectivity of the global network. The attacker's problem at time $k$ is formulated as follows:
\begin{equation}
\mathcal{Q}_A^k:\ \ \ \min_{e}\  \lambda_2(e,\mathbf{x}(k)),
\end{equation}
where $\lambda_2(e,\mathbf{x}(k))$ is the connectivity of the global network, with $e$ representing the attacker's strategy and $\mathbf{x}(k):=[\mathbf{x}_1(k);\mathbf{x}_2(k)]$ the network configuration at time $k$. On the other hand, the network operator $\gamma$'s problem is, for $\gamma\in\{1,2\}$:
\begin{equation}\label{p1}
\begin{split}
\mathcal{Q}_\gamma^k:\ \ \ &\max_{\mathbf{x}_\gamma({k+c_\gamma})}\ \min_{e}\  \lambda_2(e,\mathbf{x}({k+c_\gamma}))\\
\mathrm{s.t.}\ \ \ & \mathrm{physical\ dynamics\ of\ autonomous\ systems},
\end{split}
\end{equation}
where $\mathbf{x}_\gamma({k+c_\gamma})$ is the configuration of the mobile network controlled by operator $\gamma$ at time $k+c_\gamma$, and $c_\gamma$ is a positive integer indicating update frequency. Note that the network operator's problem falls into the general framework formulated in Section \ref{sec: DynamicGameCrossLayer}, where the dynamics of autonomous systems can be captured by \eqref{xsystem} and \eqref{ysystem}. Furthermore, each network operator needs to prepare for the worst case attacks (Stackelberg game) as well as the action taken by the other operator (Nash game) during the network reconfiguration.

This games-in-games framework has been corroborated to be effective in obtaining the self-adaptability, self-healing, and agile resilience of heterogeneous autonomous systems. In the Internet of battlefield things, the unmanned ground vehicle network coordinates its actions with the  unmanned aerial vehicle network and the soldier network to achieve a highly connected global network \cite{chen-TCNS-19-optimal}. The designed decentralized algorithm in \cite{chen-TCNS-19-games} yields an intelligent control of each agent to respond to others to optimize real-time network connectivity under adversaries. Fig. \ref{spoofing_case} shows the results of a two-layer autonomous system on the battlefield where two operators prepare for potential jamming attacks. Furthermore, the agents can respond to the spoofing attack quickly which shows the agile resilience of the control strategy.
The developed games-in-games model can be further extended to address the \textit{mosaic control design} as the framework provides built-in security and resilience for each component in the system which guarantees the performance of the integrated system.

\subsection{Multi-stage Bayesian Games: Security under Adversarial and Defensive Deception}\label{subsec:Deception}
\label{sec:Multi-stage Bayesian Game}
\begin{figure*}[htb]
    \centering
    \includegraphics[width=0.7 \textwidth]{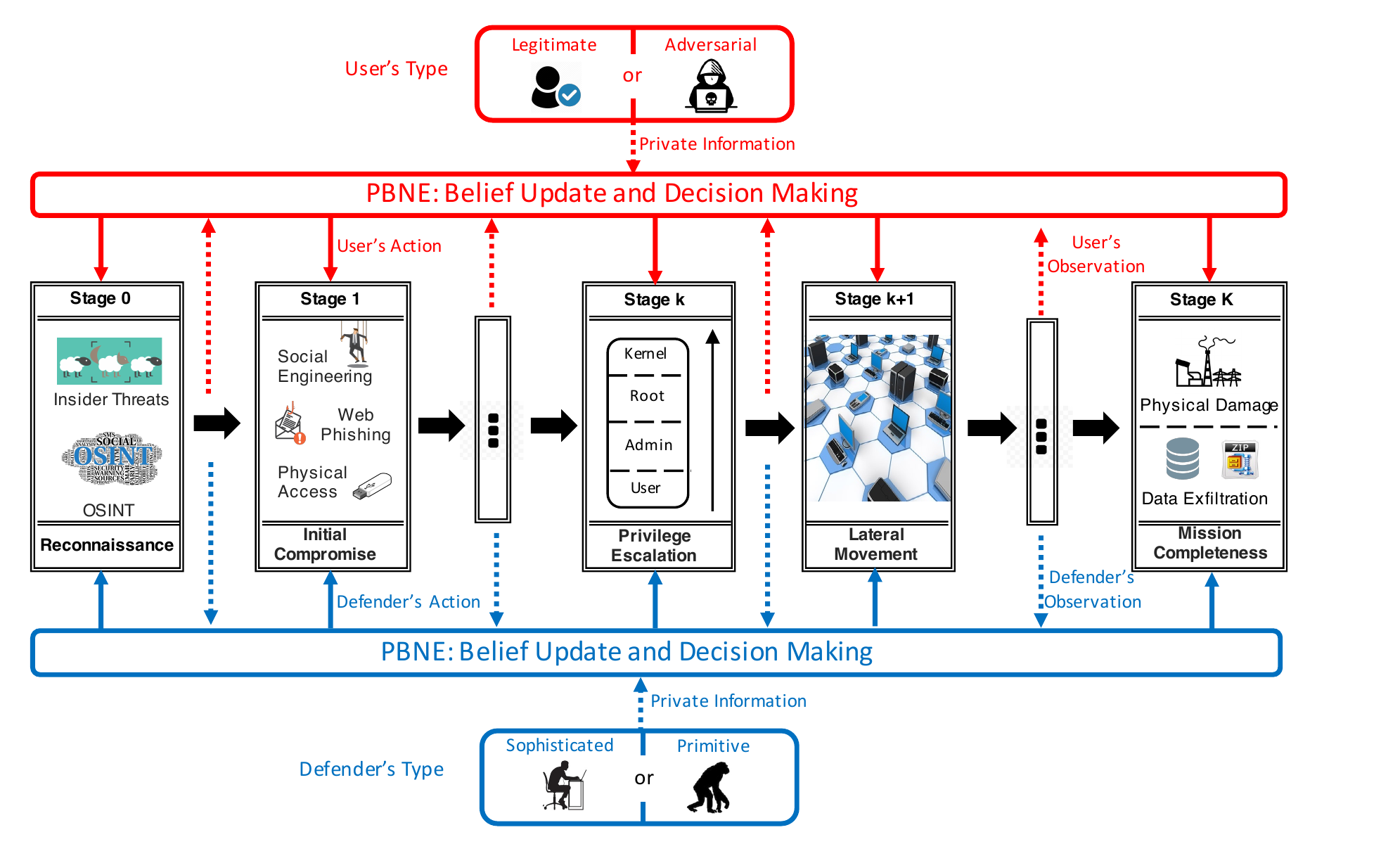}
    \caption{A block diagram of the proposed proactive defense-in-depth paradigm against multi-stage stealthy APTs. As denoted in black, each stage describes a local interaction between the user and the defender where the outcome leads to the next stage of interactions. 
    Dash arrows represent the information available to each player, which can be used to update the belief and decide the cross-stage behavioral strategy based on the PBNE. 
    Then, each player takes an action at each stage according to the strategy, as denoted in solid arrows. 
    }
    \label{StateDiagram}
\end{figure*}

APT attacks originated from a cyber network (the middle layer of Fig. \ref{fig:CrossLayerDesign}) can stealthily escalate privilege, move laterally, and lead to damage in the physical control system (the bottom layer of Fig. \ref{fig:CrossLayerDesign}). 
The entire intrusion process can be divided into multiple phases in sequence, as denoted by the black boxes in the middle layer of Fig. \ref{StateDiagram}.

Each phase serves as a stepping stone for the next phase and plays an indispensable role in the success of APTs. 
Based on the multi-stage and stealthy characteristics of APTs, 
\cite{HuangAPT} has suggested a \textit{Defense-in-Depth (DiD)} paradigm to counter them \textit{proactively}. 
DiD as the first aspect means that a control system defender should adopt defensive countermeasures at all phases of APTs and holistically consider interconnections and interdependencies among these stages. 
For example, a privilege restriction at the escalation phase can result in a failure or an additional cost for the APT attacker to take control of the targeted sensor at the final stage. 
Proactive actions and precautions as the second aspect mean that the defender needs to act before an attack is revealed.  
On one hand, these precautions can mitigate the loss induced by the APT attack at the final phase and deter attacks at their early stages. On the other hand, they can also impair the user experience and reduce the utility of legitimate users. Hence, the defender has to take judicious actions at each stage to balance usability versus security. 

The lower and upper layers of Fig. \ref{StateDiagram} illustrate a $K$-stage strategic interaction between the proactive defender and the user in blue and red, respectively. 
The type of a user can be either adversarial or legitimate. 
Since an APT attacker can pretend to be a legitimate user throughout stages, the defender does not know the user's type. The defender can observe suspicious user actions at each stage. However, these suspicious actions do not directly reveal the user's type because a legitimate user may also take them. For example, both the Tor network connection \cite{milajerdi2015composite} and the code obfuscation \cite{nissim2015detection} can be used legitimately or illegally. 
Similarly, a defender can also be classified into different types based on factors such as her level of security awareness, detection techniques she have adopted, and the completeness of her virus signature database. 
To tilt the information asymmetry that the user has a private type, the defender can also introduce defensive deception and make her type unknown to the user. 
The defender takes proactive actions at each stage and the user can observe them at the next stage. 
Therefore, each stage describes a local interaction between the attacker and the defender (two-layer game) where the outcome leads to the next stage of interactions. 
Participants receive different stage utilities from each local interaction (the discrete counterpart of \eqref{PSG}) and each player aims to  find a behavioral strategy of this dynamic game to maximize his expected utility accumulated over $K$ stages. 
The behavioral strategy means that each player needs to decide which action to take or take an action with what probability based on the available information.  
Each player introduces a belief to quantify the uncertainty of the opponent's type and adopts the Bayesian update to correlate the information revealed at each stage and reduce the type uncertainty. 
The solution concept of Perfect Bayesian Nash Equilibrium (PBNE) is introduced where `perfect' captures the cross-stage cumulative utility, `Bayesian' captures the type uncertainty, and `Nash Equilibrium' captures the strategic interaction between two players. 
The PBNE provides a creditable predication of both players' behaviors over $K$ stages because no players benefit from unilateral deviations at the equilibrium. To solve the coupling between the forward belief update and the backward equilibrium computation,  \cite{HuangAPT} has proposed a sequence of nested algorithms. 
The authors in \cite{HuangAPT} have also provided an elaborate case study of APT attacks on the Tennessee Eastman process (a specific example of \eqref{xsystem} and \eqref{ysystem}) and obtained the following insights. 

First, one ounce of proactive actions when the attack remains ``under the radar" is worth a pound of post-attack response.  
Second, the online learning capability of the defender reveals hidden information from  observable behaviors and threatens the stealthy attacker to take more conservative actions. 
Third, defensive deception introduces uncertainty to attackers, increase their learning costs, and hence reduces the probability of successful attacks. 

\subsection{Dynamic Games for Risk Management of Networked Systems}\label{subsec:RistManagement}
Game theory is widely adopted in the risk management of complex engineering systems \cite{chen_2019_TIFS,zhang2017bi}. Mitigating the risk of multi-agent systems is critical for their secure and efficient operations. However, due to complex interdependencies between nodes and the fast-evolving nature of threats, controlling the risks of multi-agent systems is not a trivial task and requires expert knowledge. Hence, one approach for the system owners is to delegate tasks of risk mitigation to security professionals, creating security as a service paradigm \cite{chen-tifs-17}. 

\begin{figure}
    \centering
    \includegraphics[width=0.8\columnwidth]{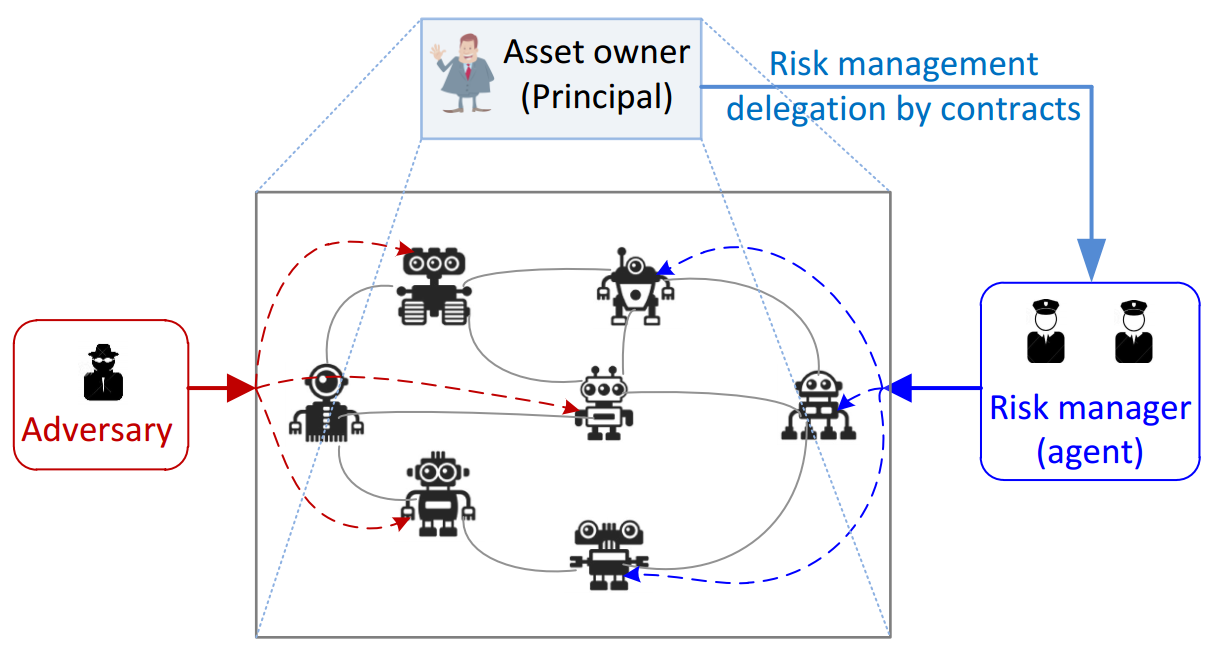}
    \caption{Risk management of a networked system through dynamic contracts. The asset owner (principal) delegates the risk management tasks to security professional (agent) by designing a contract which specifies the dynamic remuneration schemes. The agent's effort is hidden to the principal. The amount of remuneration depends on the observed risk of the system. The contract mechanism design can be formulated as a stochastic Stackelberg differential game under nonstandard information.}
    \label{risk_management}
    \vspace{2mm}
\end{figure}


As shown in Fig. \ref{risk_management}, the owner can be seen as a principal who employs a security professional to fulfill risk management tasks, and the security professional (risk manager) can be regarded as an agent whose efforts are dynamically compensated by the principal. This type of two-sided service relationship can be captured by a principal-agent framework.
One unique feature of the framework is that the principal cannot directly observe the efforts adopted by the agent. Thus, the principal needs to design a contract that specifies the compensation rules only based on observable risk outcomes. 
Specifically, the cyber risk evolution can be described by the following dynamic systems (which belongs to the general dynamics \eqref{xsystem}):
\begin{equation}\label{cyber_risk}
\begin{split}
d\mathbf{x}(t) &= f(\mathbf{x}(t),\mathbf{u}(t),t)dt + \Sigma(\mathbf{x}(t),t)d\mathbf{b}(t),\\
\mathbf{x}(0) &= x_0,
\end{split}
\end{equation}
where $f:\mathbb{R}^N\times \mathbb{R}_+^N \times [0,T]\rightarrow \mathbb{R}^N$, $\Sigma:\mathbb{R}^N \times [0,T]\rightarrow \mathbb{R}^N$ with $\mathbf{x}(t)\in\mathbb{R}^N$ represents the risk of nodes in the system, $\mathbf{u}(t)\in\mathbb{R}_+^N$ the hidden effort of the agent, $\mathbf{b}(t)$ is an $N$-dimensional standard Brownian motion, and $x_0$ is a known $N$-dimensional constant vector indicating the initial risk. The dynamic contract designed by the principal is $p(t)$, $t\in[0,T]$, reflecting the payment delivered to the agent at time $t$. First, the principal's goal is to minimize the risk $\mathbf{x}(t)$ by providing an appropriate amount of incentives $p(t)$ over the time horizon.

Second, the contract should capture the agent's behavior including the incentive compatibility (IC) and the individual rationality (IR). The principal's cost function is
\begin{equation}\label{J_P}
\begin{split}
J_P(\{p(t)\}_{0\leq t\leq T}) 
= \mathbb{E}\int_0^T f_P(t,\mathbf{x}(t),p(t))dt,
\end{split}
\end{equation}
and the agent's cost function is 
\begin{equation}\label{J_A}
\begin{split}
&J_A\left(\{\mathbf{u}(t)\}_{0\leq t\leq T};\{p(t)\}_{0\leq t\leq T}\right)\\ = &\mathbb{E}\int_0^T f_A(t,p(t),\mathbf{u}(t)) dt,
\end{split}
\end{equation}
where $\mathbb{E}$ is the expectation operator, and $f_P$ and $f_A$ are the running costs of two players. Furthermore, the IC constraint is 
$J_A\left(\{\mathbf{u}(t)^*\}_{0\leq t\leq T};\{p(t)\}_{0\leq t\leq T}\right) 
 \leq  J_A\left(\{\mathbf{u}(t)\}_{0\leq t\leq T};\{p(t)\}_{0\leq t\leq T}\right),\
  \forall \mathbf{u}(t),\ t\in[0,T],$
 and the IR constraint is 
 $\inf_{\mathbf{u}(t)} J_A\left(\{\mathbf{u}(t)\}_{0\leq t\leq T};\{p(t)\}_{0\leq t\leq T}\right) \leq \underline J_A,$
 where $\underline{J}_A$ is a predetermined non-positive constant.

This contract design for risk management problem can be formulated as a \textit{stochastic Stackelberg differential game under nonstandard information}. To design the optimal contract, the authors in \cite{chen-allerton, chen-dgaa} have developed a three-step approach including the estimation, verification and control phases, which transformed the principal's nonclassical control problem into a standard stochastic control program. For example, when the dynamics in \eqref{cyber_risk} admit a linear form and the players' cost functions are quadratic, then the optimal contract can be obtained through solving a matrix Riccati equation \cite{chen-dgaa}. Under mild conditions on the structure of cost functions of two players, the authors have revealed a \textit{separation principle} where the estimation and control phases can be addressed separately. The authors have also discovered a \textit{certainty equivalence principle} for a class of dynamic mechanism design problems where the contracts designed under incomplete case and full information scenario (the principal can directly observe the agent's action)  coincide. The contract mechanism has been corroborated effective in mitigating the risks.  

The developed framework for risk management can be applied broadly, such as industrial control systems, enterprise networks, and critical infrastructures. Furthermore, the dynamic mechanism design problem can be extended extensively, which is of great interest to the control community. For example, the underlying system could have jump parameters, the risk could be governed by mean-field dynamics in large networks, the risk cannot be directly observable to the players, and the risk observation is intermittent, etc.

\section{Conclusion and Future Development}
In this review, we have discussed recent advances and applications of dynamic games to the robust, secure, and resilient design of modern control systems. We have introduced the hierarchical structure of modern control systems, offering a holistic view of control systems that leads to an integrated dynamic game framework.  The dynamic games approach has successfully captured the multi-layer cyber, physical, and human interactions in control systems as well as their behaviors in adversarial environments. The game-theoretic modeling has provided a fundamental understanding of the tradeoffs among robustness, security, and robustness, leading to a new system science and design paradigm. 

The application of dynamic games to control systems is still in its infancy despite a rich literature in game and control theory. The bridging between these two fields would require addressing many research challenges. Computational complexity is one important research direction. Analysis of large-scale game-theoretic models is often difficult.  It would be essential to develop efficient algorithms to compose distinct models, compute equilibrium solutions, and solve mechanism design problems. These tools would lead to the core of the next-generation control system technologies that have the capabilities of automated defense, self-organizing, and fast recovery. 

Another key challenge arises from dealing with human factors at the supervisory and management layers. It has been observed that many security breaches are due to human cognitive errors, limited reasoning capabilities, and mismatched perception of risk. Integrating human modeling into control systems is critical to enable a scientific framework for human-centered design. Recent advances in behavioral game theory and epistemic game theory have laid necessary theoretical foundations for the modeling of bounded rationality and human behaviors. Hence game theory provides an unprecedented opportunity to understand human factors in control systems by bridging game theory and control system theory.

\section*{Acknowledgment}
This research is partially supported by award 2015-ST-061-CIRC01, U. S. Department of Homeland Security, awards ECCS-1847056, CNS-1544782, and SES-1541164 from National Science of Foundation (NSF), and grant W911NF-19-1-0041 from Army Research Office (ARO).

\ifCLASSOPTIONcaptionsoff
  \newpage
\fi



%

\bibliographystyle{IEEEtran}

\bibliography{ref}

%








\end{document}